# Evaluation of Time-Critical Communications for IEC 61850- Substation Network Architecture

Ahmed ALTAHER[1], Stéphane MOCANU[1], Jean-Marc THIRIET[1]
[1]Univ. Grenoble Alpes, GIPSA-Lab, F-38000 Grenoble, France
{ahmed.altaher}@gipsa-lab.grenoble-inp.fr

## Abstract

Present-day developments, in electrical power transmission and distribution, require considerations of the status quo. In other meaning, international regulations enforce increasing of reliability and reducing of environment impact, correspondingly they motivate developing of dependable systems. Power grids especially intelligent (smart grids) ones become industrial solutions that follow standardized development. The International standardization, in the field of power transmission and distribution, improve technology influences.

The rise of dedicated standards for SAS (Substation Automation Systems) communications, such as the leading International Electro-technical Commission standard IEC 61850, enforces modern technological trends in this field. Within this standard, a constraint of low ETE (End-to-End) latency should be respected, and time-critical status transmission must be achieved.

This experimental study emphasis on IEC 61850 SAS communication standard, e.g. IEC 61850 GOOSE (Generic Object Oriented Substation Events), to implement an investigational method to determine the protection communication delay. This method observes GOOSE behaviour by adopting monitoring and analysis capabilities. It is observed by using network test equipment, i.e. SPAN (Switch Port Analyser) and TAP (Test Access Point) devices, with on-the-shelf available hardware and software solutions.

## 1   Introduction

Power transmission and distribution substations are part of modern smart grids. These grids have communication services that follow recent standards [1]. In result, modern SAS communication systems are evolved according to industrial demand and convenient applicable technologies, including but not limited to switched Ethernet, VLAN and priority tagging. The IEC 61850 standard includes at least 10 parts that drive interoperability between several SAS devices, e.g. intelligent electronic devices (IEDs), from different vendors. This standard recommends use of high-speed time-critical communications for updating SAS status and events [2]. Nowadays, IEDs equipped with Ethernet NIC ports are capable of exchanging events and status to accomplish very complex functions. The standard sets a time constraint for the protection and control. A response time within less than four milliseconds is needed for high-performance class of protection (protective relay) requirements [4].

Aiming to evaluate reliability of station bus IEC 61850-8-1 [1] communications, GOOSE is selected as messaging service for the experimental setup. The objectives are to observe the behaviour of the SAS functional aspects, particularly, for events and status exchange, and to find the ETE delay between understudy IEDs that participate in the protection functionalities of primary equipment (process level).

To obtain a real-time traffic and to determine ETE delay with actual SAS devices, a distribution SAS platform, equipped with industrial control devices, is installed and configured. This platform includes several automation devices from different vendors. The SAS platform is used for research purpose as well as for pedagogical tasks. A network with industrial switches was designed to represent star topology network that is latterly implemented to provide 100 Mbps full duplex bandwidth.

This paper is organized as the follows: First, section 2 provides an overview about the literature review. Section 3 presents the switched Ethernet based station-bus. Additionally, section 4 introduces the IEC 61850 with more details in its subsections. Afterwards, section 5 shows the experimental platform, and section 6 explains the experimental setup. Finally, section 7 discusses the results obtained.

## 2 Literature review

According to the IEC 61850-8-1 documentation, GOOSE messages are reliable for event and status exchange between IEC 61850 based substation automation devices been either for power protection or for control. Some applications such as circuit breaker closure, current differential parameters acquisition utilize IEC 61850-8-1 communication service for event and status delivery within real time protection and control facilities. To perform real-time delivery GOOSE communications offer time-critical feature by using Ethernet (IEEE 802.3) multicasting paradigm.

In other hand, performance specification for SAS critical-time communication requires less than four ms ETE Application-to-Application latency. Researchers studied this constraint from different aspects.

Modelling and simulation approaches were pursued to simulate communications of both Ethernet based GOOSE and SV (sampled values). Other researchers addressed co-simulation methodology, i.e. HITL (Hardware in the Loop), for studying SAS communication behaviour, they adopted SAS functions without using complete interoperable devices. These co-simulations neglected some parameters such as different GOOSE frame size, number of IEDs, and media type.

Therefore, to the best of our knowledge, all previous efforts were either to calculate the delay theoretically, or to find the delay by modelling the SAS behaviour, and by using simulation software. In [5] Network Calculus Algebra was used to determine worst-case boundaries for SV (Sampled value) and GOOSE traffic at bay level, but with estimated parameters for network traffic. Modelling and simulation approach was conducted using OPNET Modeller In [6], where a feeder bay model was created and priority scheme was adopted. The previous mentioned approach also was used in [7] to build a distribution SAS star-topology network with IEDs models for MU (Merging Unit), Protection and CB (Circuit Breaker).

In our research, we focus on real-time experimental platform dedicated for research purposes. The platform is built upon industrial standards with cyber-physical components to implement substation automation functions that include communication, monitoring, protection and control.

## 3 Switched Ethernet based station-bus

Communication networks become a part of SAS networked control systems. Distributed processing and control benefit from networking capabilities in digital SAS systems, thus, Ethernet based technology is suitable for local communications in the field of power protection and automation. The original Ethernet was released in 1976 and its technologies have been in use since 1979. Local area networks (LANs) are built highly upon Ethernet technologies to allow high bandwidth and low cost. IEEE 802.3 is the key standard for Ethernet LANs, in early days, Ethernet uses a carrier sense multiple access with collision detection (CSMA/CD) methodology for media control, i.e. transmission and collision management, and linking of network nodes. Ethernet utilise datagram (frames) to send data units including source and destination address, payload data, and error control. Its bandwidth uses different media (cables) such as coax, twisted pair and optical fibre. Thereafter, it evolved during four generations: standard 10 Mbps, fast 100 Mbps, Giga bit Ethernet with one Gbps, ten-Giga bit with 10 Gbps [8]. First step of Ethernet evolution was using bridging concept in order to increase the bandwidth and to separate collision domains [8]. Switched Ethernet is extension of bridged LAN to eliminate collisions and to allow faster Ethernet and full-duplex transmission.

The need for higher performance and real-time data exchange over Ethernet enforces using of protocols that would be suited to TCP/IP protocol (Transmission Control Protocol/ Internet Protocol). Among these protocols are Modbus over TCP/IP protocol data unit (PDU), and Distributed Network Protocol (DNP 3.0) that has been implemented in similar manner as Modbus. Thus, these protocols uses Ethernet layer. The IEC standards group provided a standard (IEC 870-5-104) that outlines how to encapsulate the IEC 870-5-101 over Ethernet. Hence that, the IEC 61850 standard allowed using of Ethernet technologies for GOOSE, SV and MMS communications [9].

Switched Ethernet is evolved to allow priority tagging (IEEE 802.1p), i.e. classification of transmitted frames, to enable different class of service including higher priorities for critical applications. In addition, the IEEE 802.1Q introduces virtual LANs (VLAN) concept to reduce collision domains, by only transmitting frames to target VLAN [10]. These techniques added fields for Ethernet frames (fig. 1)

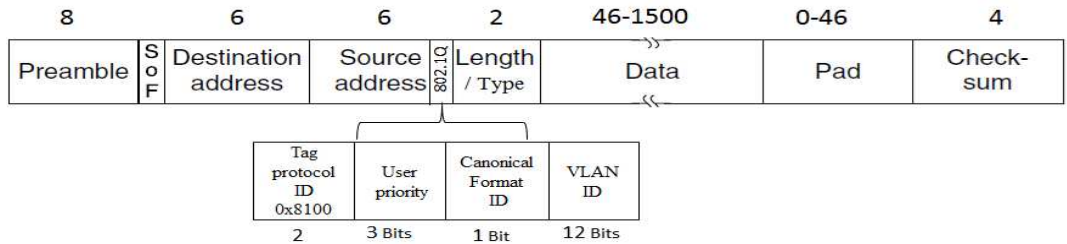

Figure 1: Ethernet frame with VLAN tagging and priority (IEEE 802.1p/q)

## 4  IEC 61850

The Technical Committee 57 (TC57) belongs to the International Electro-technical Commission (IEC) has released the IEC 61850 standard that enforces interoperability between SAS devices, e.g. IEDs, and enables abstraction of communication services [2]. It provides communication services standards for station level, bay level and process level equipment as demonstrated in (fig.2).

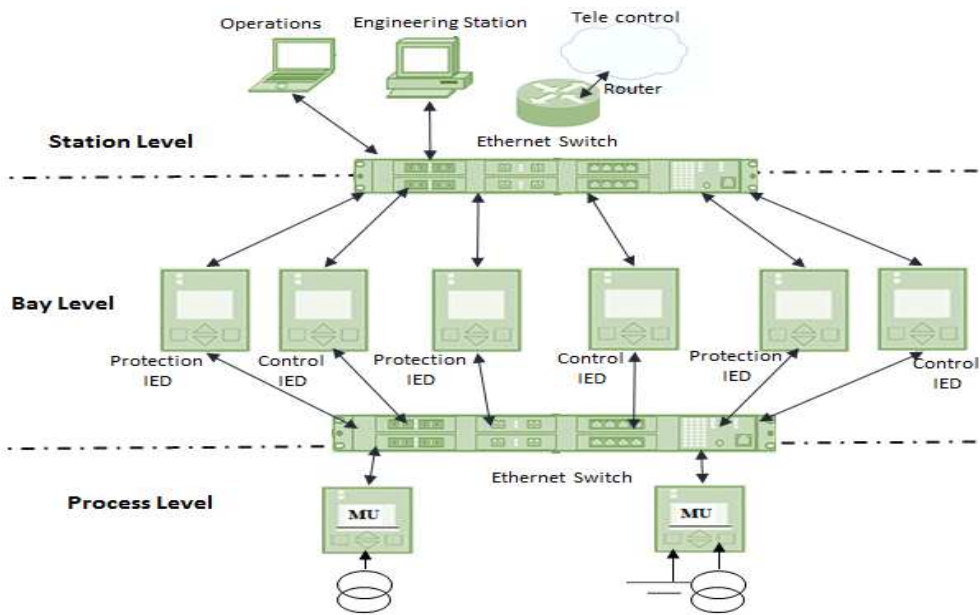

Figure 2: IEC61850 Substation Automation System

The standard defines the various aspects of substation communication system in 10 parts; first five parts define and identify the general and specific functional requirements for SAS communications. In part 6, a configuration description language called Substation Configuration Language (SCL) gives capability of SAS configuration as engineering tool; it allows the description of the relations between the substation automation system and the substation equipment (switchyard) [3]. Part 7 includes four subsections dealing with abstraction and mapping of data services to communication protocols, which means representing the physical components by logical nodes. Data and services abstraction gives a step toward mapping these objects into actual protocols. Part 8 defines this mapping into Manufacturing Message Services (MMS), while part 9 including two sections to define the mapping of sampled measurement values (SV) in section 9-1 and the process bus in section 9-2. The last part, part 10 gives a detailed approach for conformance testing.

## 4.1 IEC 61850 communications reliability

Reliability is defined as trustworthiness of a system with respect to its continuous delivery of correct service or, equivalently, the time to failure [11]. The IEC 61850 part 3 section 4 considers reliability as quality requirement [2] by focusing on communications for SAS networking services. In addition, the standard gives a reference for other standards such as IEC 60870-4 that specifies performance requirements for tele-control. Further, IEC 61850 defines SAS communication reliability as continuation of SAS data networking without failure. Precisely, there should be no single point of failure in SAS networks. If failure exist, outcomes may cause damage to SAS equipment. The standard insists that communication reliability is needed as a requirement for SAS functional implementations; therefore, it recommends fail-safe design that should be handled to avoid undesired control action.

## 4.2 IEC 61850 GOOSE

The Generic Object Oriented Substation Event is defined in the IEC61850 part 7-1. This message is classified within type1 performance that means high-speed message to deliver status and event changes. Among these classified messages are type 1-A which is mission critical that require less than four milliseconds ETE delay. The built-in reliability for GOOSE come from retransmission of message frame until end of its life according to predefined field named Time Allowed To Live (TATL property). The GOOSE transmission period is determined by minimum and maximum retransmission period. The maximum frame size equals maximum Ethernet frame with IEEE 802.1p/q encapsulation (fig. 1), i.e. Ethernet size with priority tagging and VLAN fields equals 1822 bytes. Some factors will affect GOOSE communication delay, among of them, switch fabric and switching technology, communication media, available bandwidth, traffic pattern, etc. GOOSE service would benefit from Ethernet priority tagging capability to improve time-critical priority demands. The standard recommends GOOSE priority will be greater or equal to level four (maximum is seven). Actually, GOOSE frames have an assigned Ethernet-type (fig. 1) that equals hexadecimal number 0x88b8. Network analysers recognize GOOSE frames by their Ethernet type field.

## 4.3 IEC 61850 Sampled Value

SV carries analogue measurements values and it transmits instrumentation measurements, embedded into multicast Ethernet frames, such as current and voltage to SAS devices. A good example for this communication service is to send data with a certain sample rate such as 80 responses. If the sample rate equal to 256 samples per cycle and the frequency is 50 Hz, samples obtained will be 12800 samples. The sampled value frame contains APDU (Application Protocol Data Unit) that consequently contain ASDU (Application Service Data Unit) which carries four sampled values for three phases and neutral wire. Knowing that frame size, one can calculate the produced traffic by multiplying frame size by number of samples. In result, this sampling produces 23.55 Mbps of network traffic [5].

## 5 GICS Platform Description

GICS, GreEn-ER (Grenoble Energy Teaching and Research) Industrial Control System, is a platform for learning and research activities. The GICS platform is an industrial system laboratory that includes control, e.g. programming logic controllers PLCs, and cybersecurity subsystems, e.g. protocol analysers. This platform has industrial devices representing substation automation functional services with engineering workstations. It includes three levels of SAS communication services, offering interoperability-testing environment, that contain station level workstations, bay level devices and process level equipment from different vendors. Its real-time communications include fieldbus, Modbus, IEC 61850 with variety of communication services. In this platform, power protection and control IEDs were deployed in a bay level to exchange status and control, receive commands from station level workstations, and to gather events, status and physical measurement from process bus equipment. In the process level, developed embedded cards are installed to acquire current and voltage metering from instrumentations. The GICS platform communications contain network architecture (fig.3) used for supervision, protection and control, and monitoring. For detailed analysis of protection and control communications, the GICS devices transmit network traffic including Ethernet based IEC 61850 frames and TCP/IP packets.

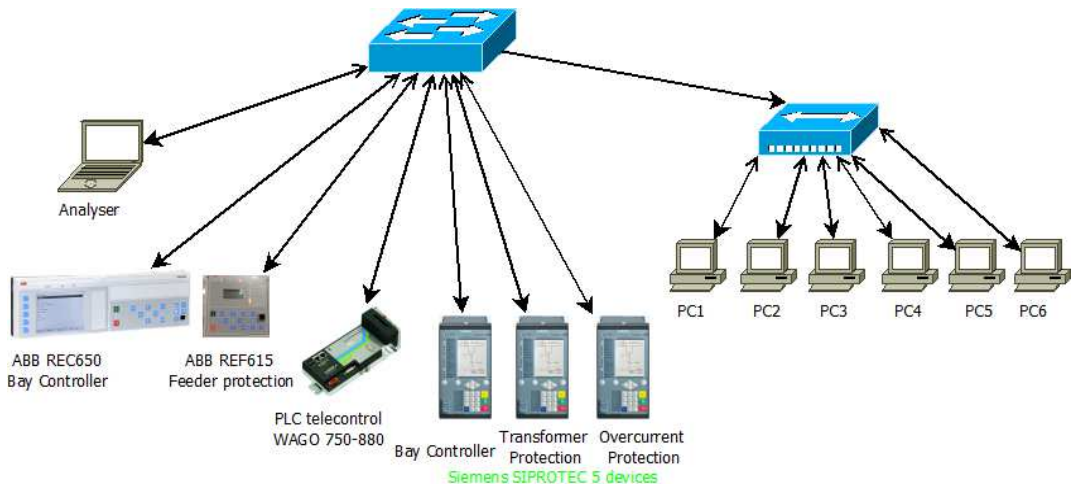

Figure 3: GICS platform network architecture with SAS devices.

## 6 The Experimental Setup

To capture the traffic and to get frames at different points network access devices were connected at both publisher IED and subscribed IED. Two network TAPs (Test Access Points) were used to sniff network traffic, the first network TAP was installed to access network traffic generated by the publisher IED, while the other TAP was installed to access network traffic at the subscribed IED.
An analyser laptop, equipped with two network interfaces and analyser tool, is used to capture network traffic from both TAP nodes. After gathering the traffic, the captured frames are filtered according to: 1) a specified communication protocol, which is GOOSE, and 2) physical address (MAC address) of publisher IED. We saved the captured frames in a file for later analysis and processing (fig.4).

Figure 4: GOOSE captured frames

Captured GOOSE frame (fig. 5) is distinguished by the sequential number field (SqNum) and timestamp (t). The frames contain fields according to the IEC 61850 standard.

Figure 5: GOOSE sequential number

The ETE delay is determined by subtracting timestamps of the successive publisher and subscriber frames, that has the same sequential number but different timestamps. To calculate the delay, the following formula is used:

$$T_{delay} = T_{des} - T_{tr} \qquad (1)$$

Where $T_{delay}$ is delay time, $T_{des}$ is timestamp at subscriber IED and $T_{tr}$ is timestamp at publisher IED. Delta time (latency), for captured frames, is measured within microsecond precision. Finally, the average delay is figured according to the number of captured frames during the experiment period. This average value is calculated by dividing overall delay (sum of frames delay) by number of frames.

# 7 Results and discussion

The analyser computer captured all the network traffic through switch port analysed node (SPAN). Likewise, two TAPs (Test Access Points) captured publisher and subscriber IEDs traffic with zero-loss and zero-delay. We observed all network load during a period of three minutes (fig. 6).

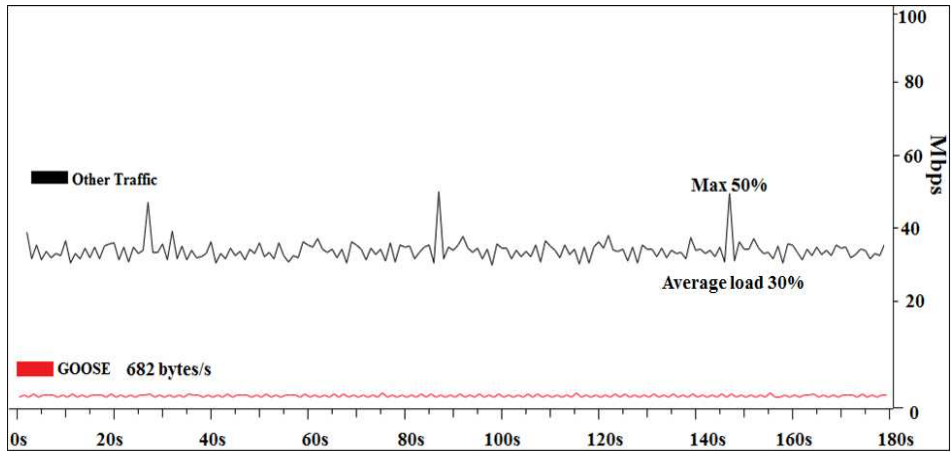

Figure 6: Network traffic load with GOOSE frames

The captured traffic included frames timestamps that used latterly to calculate the ETE delay. The network traffic is filtered retrieving messages frames for only four devices (table 1) producing GOOSE communications (fig. 7).

| Device | Function | GOOSE size |
|---|---|---|
| Vendor 1 device 1 | Overcurrent Protection | 162 Bytes |
| Vendor 1 device 2 | Transformer Protection | 150 Bytes |
| Vendor 2 device | Feeder Protection | 178 Bytes |
| Vendor 3 device | Tele-control PLC | 192 Bytes |

Table 1: GOOSE based communicated devices

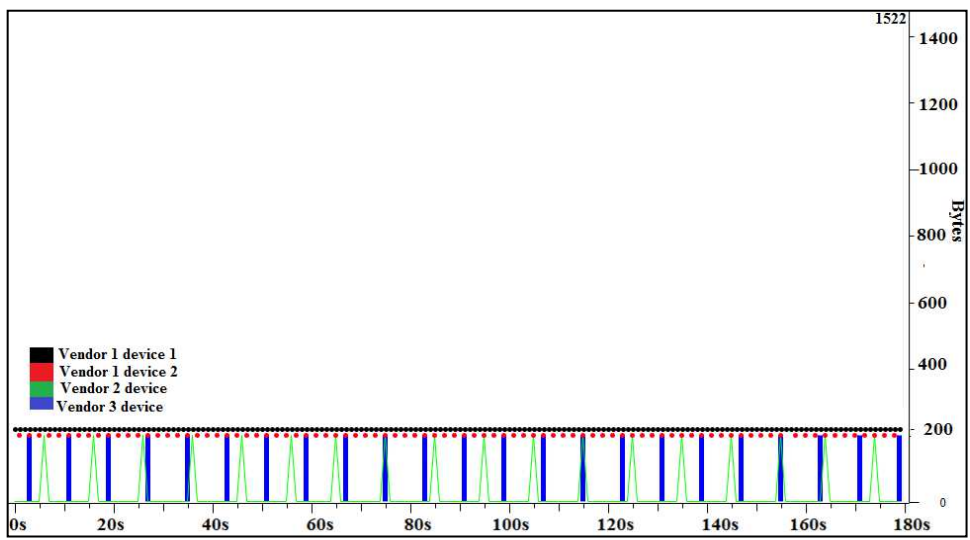

Figure 7: Network traffic load with GOOSE frames

The delay average is determined equal to 26 µs when the average load is 30%, this value represents adequate latency respects the time-critical performance requirement. Adding more traffic reaching 50% load saturation of communication channel scenario, 310 µs latency is obtained. Also adding SV generated traffic during GOOSE frame retransmission causes a delay equal to 1.1 ms. From the mentioned results we found that 100 Mbps Ethernet network with VLAN tagging and priority configuration is suitable for GOOSE communication that serve critical-time performance.

# 8 Conclusion and Perspective

We observed the IEC 61850-network traffic during the experimental activities. Real-time GOOSE communication behaviour is analysed to determine the ETE delay of publisher/subscriber mechanism. The results obtained provided suitable delay average respecting critical-time performance for the protection and control requirements. In future work, the researches intended to investigate station-bus network reliability by adding communication redundancy for station-level architecture. For reliability investigation, redundant switch would be deployed with using of redundant network interfaces in every IED. This work would include determining the recovery time and its effect on ETE delay.